\documentclass[aps,twocolumn,showpacs,floats,floatfix,prb]{revtex4}
\usepackage{epsf}
\usepackage{float}
\usepackage{amsmath}
\usepackage{graphicx}

\begin{document}

\title{Theory of a resonant level coupled to several
conduction electron channels in equilibrium and out-of-equilibrium}

\author{L\'aszl\'o Borda$^1$}
\author{K\'aroly Vlad\'ar$^2$}
\author{Alfr\'ed Zawadowski$^{1,2}$}
\affiliation{
$^1$Department of Theoretical Physics
and 
Research Group ``Theory of Condensed Matter'' of the Hungarian Academy of
Sciences,\\ 
Budapest University of Technology and Economics,
Budafoki \'ut 8.  H-1521 Budapest, Hungary\\
$^2$Research Institute for Solid State Physics and Optics, 
P.O. Box 49, H-1525 Budapest, Hungary}

\begin{abstract}
The spinless resonant level model is studied when it is coupled by
hopping to one of the arbitrary number of conduction electron
channels. The Coulomb interaction acts between the electron on the
impurity and in the different channels. In case of repulsive or
attractive interaction the conduction electrons are pushed away or
attracted to ease or hinder the hopping by creating unoccupied or
occupied states, respectively. In the screening of the hopping
orthogonality catastrophe plays an important role. In equilibrium
the weak and strong coupling limit the renormalizations are treated
by perturbative, numerical and Anderson-Yuval Coulomb gas methods.
In case of two leads the current due to applied voltage is treated in
the weak coupling limit. The presented detailed study should help to
test other methods suggested for non-equilibrium transport.
\end{abstract}
\date{\today}
\maketitle

\section{Introduction}

In the recent years the quantum impurity problem out-of-equilibrium has
attracted great interest. The most relevant realizations are the 
quantum dots connected to at least two metallic leads\cite{qdot} 
and
short metallic wires
containing magnetic impurities\cite{wire}. 
In the impurity problem the exact methods 
play
distinguished 
roles
especially the Bethe Ansatz and conformal
invariance. The generalization of these methods to out-of-equilibrium
situations are the most challenging new  
directions. Mehta and Andrei are aiming to
solve the Kondo problem on a dot with two contact attached. First 
a simple resonant level without spin was studied
to test the new generalization of the Bethe Ansatz method\cite{mehta}.
Their elegant suggestion is very provocative. In order to 
test
this kind of new methods we perform a detailed study of that problem using
different weak coupling perturbative methods combined with NRG. As the final
goal we calculate the current flowing through the impurity when a finite
voltage is applied on the contacts. The most challenging claim of Mehta and
Andrei is that the current 
is a non-monotonic function of the strength of the Coulomb coupling
between 
the electron on the dot and conduction electrons in the two leads. 

In order to make the comparison more explicite we generalize the
time-ordered scattering formalism for non-equilibrium in the next leading
logarithmic order. In this way the current is calculated as a function of the
applied voltage and the Coulomb coupling strength. 
Increasing the Coulomb coupling strength
we find 
also
a non-monotonic
feature  but 
the order of increasing and decreasing regions is the opposite to the
finding of Mehta and Andrei\cite{mehta}.

The model to be treated is the following:
A single impurity orbital is coupled
to two reservoirs of Fermi gas via hopping but the two reservoirs have 
different chemical potentials $\mu_L$ and $\mu_R$ 
on
left and right of 
the impurity in a one dimensional model.
$\mu_L-\mu_R=eV$ is determined by the applied voltage $V$ ($e$ is the 
electronic charge). The Coulomb interaction acts between the electron on the
impurity level and the conduction electrons at the impurity position.
Thus the Hamiltonian has the form
\begin{equation}
H=H_0+H_1+H_2\;,
\label{Hs}
\end{equation}
with
\begin{equation}
H_0=\sum_{\genfrac{}{}{0pt}{}{\alpha=L,R}{k}}(k-k_\alpha)v_Fa^\dagger_{k\alpha}a_{k\alpha}+
\varepsilon_dd^\dagger d\;,
\label{H_0}
\end{equation}
where $k>0$ and $k_L-k_R=eV/v_F$, $v_F$ is the Fermi velocity,
$a^\dagger_{k\alpha}$ is the creation operator of the spinless Fermion
in lead $\alpha=L/R$, while $\varepsilon_d$ is the energy of the local level and
$d^\dagger$ is the creation operator for the electron on that site.
The interaction term is
\begin{equation}
H_1=U(d^\dagger d-\frac{1}{2})\left(\sum_{\alpha=L,R}
a^\dagger_{\alpha}a_{\alpha}-\frac{1}{2}\right)\;,
\label{H_1}
\end{equation} 
where $U$ is the Coulomb 
coupling
which in a physical case $U>0$,
$a_\alpha=\frac{1}{\sqrt{L}}\sum_ka_{k\alpha}$, and $L$ is the length of the
chain. The existence of the substraction of $1/2$ is not essential, they
can be omitted and than $\varepsilon_d$ is shifted as 
$\varepsilon_d-U/2$ and a local potential $-\frac{1}{2}U$ is acting on the
electrons, but 
the 
latter one
can be taken into account by changing the electron density
of states in the leads at the position of the impurity.

The hybridization between the lead electrons and the localized electron is
described by
\begin{equation}
H_2=V_\alpha\sum_\alpha\left(d^\dagger a_\alpha+a^\dagger_\alpha d\right)\;,
\label{H_2}
\end{equation}
where $V_\alpha$ is the hybridization matrix element.

In case of equilibrium it is useful to generalize the model to $N$ reservoirs
instead of $L$, $R$, and then $\alpha$ runs through $\alpha=0,1,\dots,N-1$
and $\mu_\alpha=\mu$. Then the hybridization term in $H_2$ is chosen in
a specific form
\begin{equation}
H_2=V_0\left(d^\dagger a_0+a^\dagger_0 d\right)\;,
\label{H_2a}
\end{equation}
indicating that only the electrons with $\alpha=0$ are hybridizing while
the others are taking part only in the Coulomb screening. Namely, only those
electrons are hybridizing which have the symmetry of the localized
orbital ($s$-like). As a result of the screening the electron gas is polarized
depending on the occupation of the localized state and that polarizations lead 
to orthogonality catastrophe\cite{anderson1}.

The model with $N=1$ is known as a resonant level model and has been studied
in great detail\cite{resonant1,schlottmann} 
and the one with $N\geq1$ has been introduced 
to study
finite range 
interaction in 3D\cite{3dmodel}.  

The goal of the present paper is to provide weak coupling results for
$V\neq0$. But before doing that the $V=0$ equilibrium case is studied 
in the weak coupling limit by diagram technique and 
then to extend the results for stronger couplings
Wilson's numerical
renormalization group (NRG)\cite{NRG_ref} 
and Anderson-Yuval Coulomb gas 
method\cite{yuval}
in order to check 
the 
validity of
weak coupling
results concerning a specific behavior. Namely at some stage of the
calculation in the exponent of the renormalized quantities a combination

\begin{equation}
-\varrho_0 U+\frac{1}{2}N(\varrho_0 U)^2
\label{exponent}
\end{equation}
appears. For $U>0$ that is changing sign at $\varrho_0 U=\frac{2}{N}$ and that 
leads in changing the increasing to 
decreasing behavior but that crossover is well
beyond the validity of the perturbation theory at least for
$N=2$.

In order to judge the real situation, an NRG study will be performed including
the weak ($\varrho_0 U\ll1$) as well as strong coupling regions 
($\varrho_0 U\geq\frac{2}{N}$) to get an insight whether the crossover indicated
above is expected or it is likely artifact 
of
the weak coupling theory.
We also map the problem to a one-dimensional Coulomb model closely 
following the work of Anderson and Yuval, where the screening can even be in
the strong coupling limit. All of these methods are suggesting 
a coherent picture of
the crossover
and they agree very well especially for $N=4$.

The study of such a crossover is especially relevant as in the work of
Mehta and Andrei\cite{mehta} such a crossover is suggested in the current
flowing in the non-equilibrium case $V\neq0$ at $\varrho_0 U\sim2$.
If we could find the crossover already in equilibrium then it is obvious 
to expect that in the non-equilibrium situation. 

The paper is organized in the following way:
In Section~\ref{sec:weak_coup}
we provide the analytical perturbative method up to next to leading
logarithmic order introducing extra channels for screening, where the
non-monotonic competion of the vertex and self-energy correction is already
demonstrated in equilibrium.
In Section~\ref{sec:nrg}
the equilibrium calculation is extended to strong coupling by using
Wilson's numerical renormalization group technique and the result is
compared to that of the analytical calculation.
In Section~\ref{app:karcsi} the Anderson-Yuval method is 
presented.
In Section~\ref{sec:out_of_equ} 
the time dependent scattering method is applied for non-equilibrium
closely following the generalized version of Anderson's poor 
man's scaling in the next to leading order and the current is calculated.
In Section~\ref{sec:summary} the results are summarized. In 
Appendix A some cancellation due to Ward indetities are discussed.

\section{Perturbation theory: weak coupling limit}
\label{sec:weak_coup}
The resonant level model is given by
Eqs. (\ref{Hs}), (\ref{H_0}) and (\ref{H_2}).
That does not contain non-commuting terms, thus Kondo behavior
is not expected in the weak coupling limit. In the strong coupling
limit that model, however, can be mapped to an anisotropic 
Kondo model\cite{resonant1,schlottmann,3dmodel}
but that mapping is not considered here. The model shows strong 
similarities to the X-ray absorption\cite{roulet,nozieres}, 
as the strength of the
interaction (invariant charge) between conduction electrons
and the electron on the impurity level is scale invariant.
The system shows scaling in terms of reduction of the conduction 
electron band width, $D$. In case $N=1$ the scaling equations
were derived by Schlottmann\cite{schlottmann} and those can be 
easily extended for arbitrary $N$.

\begin{figure}
\includegraphics[width=0.9\columnwidth]{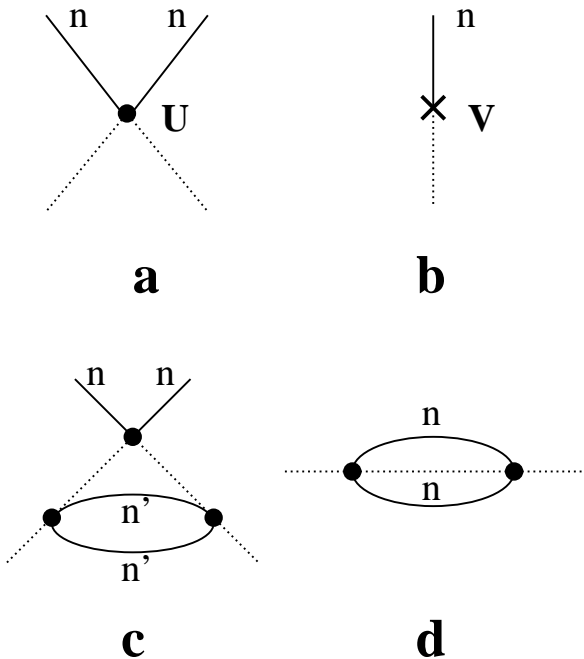}
\caption{
Vertex diagrams (a,b,c) and the impurity self energy (d).
Solid lines stand for 
conduction electron
propagators
while the dotted line for those of electron on the 
impurity level. The interactions are indicated by dots ($U$) and
crosses ($V$). In case of conduction electrons the channel
indices are also indicated.
}
\label{fig:vertex_corr}
\end{figure}

There are two different kinds of vertex corrections depicted in 
Fig.\ref{fig:vertex_corr} a,b,c, where the solid lines stand for 
conduction electrons, the dotted line for electron on the 
impurity level and the interactions are indicated by dots ($U$) and
crosses ($V$). In case of conduction electrons the channel
indices are also indicated. The Hartree-Fock energy shift
can be 
incorporated by
$\varepsilon_d$. 
The self-energy 
of the electron on the impurity is depicted
in Fig.\ref{fig:vertex_corr} d.   
In the calculation of the self-energy counter terms are introduced
to eliminate the constant terms to keep $\varepsilon_d=0$ unrenormalized.
Closely following the earlier works \cite{schlottmann,fowler,solyom}, the
invariant charge for the Coulomb interaction takes the form
\begin{equation}
U_{\rm inv}=\Gamma(\omega/D)d(\omega/D)\;,
\label{eq:inv_charge}
\end{equation}
where $\Gamma$ is the vertex function and 
$$d(\omega/D)=G(\omega,D)(\omega-\varepsilon_d)$$ can be determined
perturbatively starting with 1
where $G$ is the 
renormalized
one-electron Green's function. 
The functions $\Gamma(\omega)$ and
$d(\omega)$ are
\begin{eqnarray}
\Gamma(\omega)&=&1+N\varrho_0^2U^2\log(\frac{D}{\omega})+\dots\nonumber\\
d(\omega)&=&1-N\varrho_0^2U^2\log(\frac{D}{\omega})+\dots\;,
\label{eq:Gamma_d}
\end{eqnarray}
where $\varrho_0$ is the conduction electron density of states, for spinless
electrons in one of the channels $n=1,\dots,N-1$. As the Coulomb interaction
is independent of $n$, thus the factor $N$ occurs. As the 
consequence of the Ward identity relating the vertex correction and the
self-energy depicted in Fig.\ref{fig:vertex_corr}c and d cancel out
in $U_{\rm inv}$\cite{schlottmann,roulet,nozieres,solyom} 
\begin{equation}
\frac{d}{d|\omega|}\left(\Gamma(\omega)d(\omega)\right)=
\frac{d}{d|\omega|}U_{\rm inv}(\omega)=0\;,
\label{eq:uinv_noscaling}
\end{equation}
thus $U_{\rm inv}=U$. The renormalization group gives
\begin{equation}
d(\omega)=\left(\frac{\omega}{D}\right)^{N\varrho_0^2 U^2}
\label{eq:d_scaling}
\end{equation}
for arbitrary $n$.

\begin{figure}
\includegraphics[width=0.6\columnwidth]{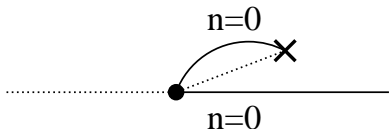}
\caption{
Vertex correction to the hybridization. 
}
\label{fig:v_corr}
\end{figure}

The hybridization contains vertex correction for $n=0$ (Fig.\ref{fig:v_corr})
\begin{equation}
V(\omega)=V\left(1+U\varrho_0\log\frac{D}{\omega}+\dots\right)\;,
\label{eq:v_corr}
\end{equation}
but it does not contain linear contribution in $\log(D/\omega)$ in the
order of $U^2$. The relevant invariant charge $V_{\rm inv}(\omega)$ is
\begin{equation}
V_{\rm inv}(\omega)=V(\omega)d^{1/2}(\omega)
\label{eq:v_inv}
\end{equation}
as the interaction is connected by only one impurity line. Thus the 
terms linear in $\log(D/\omega)$ are
\begin{widetext}
\begin{equation}
V_{\rm inv}(\omega)=V\left(1+\varrho_0 U\log\frac{D}{\omega}-
\frac{1}{2}N(\varrho_0 U)^2\log\frac{D}{\omega}+\dots\right)\;.
\label{eq:v_corr2}
\end{equation}
\end{widetext}
The result of the renormalization equation is 
\begin{equation}
V_{\rm inv}(\omega)=V\left(\frac{\omega}{D}\right)^{-\varrho_0 U+\frac{1}{2}N\varrho_0^2U^2}\;.
\label{eq:v_scaling}
\end{equation}
The second term in the exponent appears as reduction of $V_{\rm inv}(\omega)$
describing the Coulomb screening in the $N$ channels. 

For $U<0$ the $V_{\rm inv}(\omega)$ interaction is always decreasing as 
$\omega$ is reduced but for $U>0$ it depends on the strength of $U$.
For large enough $U$ the screening dominates thus
\begin{equation}
V_{\rm inv}(\omega)=
\left\{
\begin{array}{lll}
{\rm decreasing} & {\rm for} & U\varrho_0<0\\ 
{\rm increasing} & {\rm for} & 0<U\varrho_0<\frac{2}{N}\\
{\rm decreasing} & {\rm for} & U\varrho_0>\frac{2}{N}
\end{array}
\right.\;.
\label{eq:V_scaling_summary}
\end{equation}
That behavior will
be further discussed in Sec.\ref{sec:summary}.

The scaling regions, however, are not unlimited as the impurity 
level has its own
width, $\Gamma_{\rm imp}$. There are two contributions to the
level width $\Gamma_{\rm imp}$.

The hybridization 
broadens the impurity level just in case of the
Anderson model and that is in the second order in $V$
\begin{equation}
\Gamma_{\rm imp}(\omega)=\pi\varrho_0V^2(\omega)=
\pi\varrho_0V\left(\frac{\omega}{D}\right)^{-2\varrho_0 U+N(\varrho_0 U)^2}\;, 
\label{eq:gamma}
\end{equation}
where also the effect of renormalization is taken into account. There is also
a Korringa like broadening due to the creation of electron-hole pairs thus
$\Gamma_{\rm Korringa}\sim U^2\omega$, where $\omega$ comes from the phase
space. 
That is important only 
for large $\omega$ where everything is smooth thus the broadening is not
effective.

The broadening due to the hybridization cuts off the renormalization procedure 
at energy $\omega\sim\Gamma_{\rm imp}(\omega)$. Combining
Eqs.(\ref{eq:v_scaling}) and (\ref{eq:gamma}) and inserting the condition given
above 
provides
the final $V_{\rm ren}$ value as
\begin{equation}
\frac{V_{\rm ren}}{V}=\left(\frac{V^2\varrho_0\pi}{D}\right)^{\frac{-\varrho_0 U+\frac{1}{2}N(\varrho_0 U)^2}{1+2\varrho_0 U-N(\varrho_0 U)^2}}\;.
\label{eq:v_ren}
\end{equation}

\begin{figure}
\includegraphics[width=0.9\columnwidth]{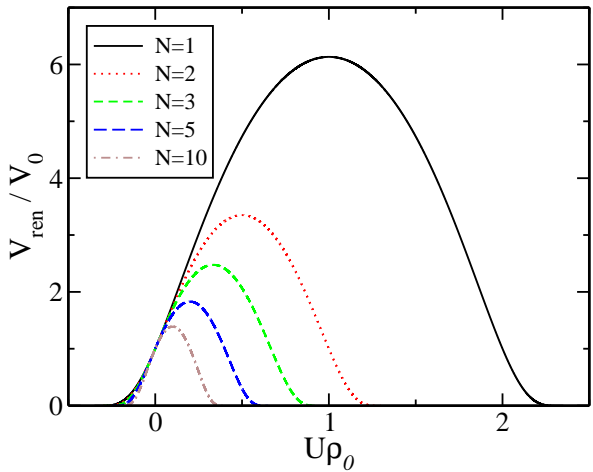}
\vspace{0.5cm}
\caption{(Color online)
The renormalized hybridization as a function of $\varrho_0 U$ for 
different channel numbers. The intermediate maximum
can diverge only for $\varrho_0 U=1$, in all the other cases the increases are
also rather moderate.
}
\label{fig:v_ren}
\end{figure}

As shown in Fig.\ref{fig:v_ren},
for $U<0$ $V_{\rm ren}/V<1$ renormalizes downwards 
and for even more negative $U$ down 
to zero.
For $U>0$ and $\varrho_0 U<2/N$ first $V_{\rm ren}$ increases with increasing
$U$ but for $\varrho_0 U>2/N$ it starts to decrease and tends to zero again. 
The intermediate maximum
appears at $\varrho_0 U=1/N$. For $N=2$
that maximum is, however, already outside the weak
coupling limit, where the calculation cannot be trusted.  The
question 
still remains
unanswered whether the non-monotonic behavior for $U>0$ can be traced
in strong coupling calculation or not. The conclusions for the crossover 
might
be trusted only for large $N\gg1$, which does not have physical relevance. 

\section{NRG approach for $V=0$}
\label{sec:nrg}
In order to determine the region of validity of the weak coupling approach
in equilibrium, we have performed numerical renormalization 
group\cite{NRG_ref} 
(NRG) analysis for the $N=2$ and $N=4$ cases. 

In Wilson's NRG technique
  --after the logarithmic
discretization of the conduction band-- one maps the original
Hamiltonian of an impurity problem to a semi-infinite chain with the
impurity at the end of the chain. As a consequence of the logarithmic
discretization the hopping amplitude along the chain decreases
exponentially as $t_n\sim\Lambda^{-n/2}$ where $\Lambda>1$ is a
discretization parameter (we have used $\Lambda=2$ throughout the
calculations) while $n$ is the site index. The separation of energy
scales provided by the exponentially vanishing hopping amplitude
allows us to diagonalize the Hamiltonian iteratively to approximate
the ground state and the excitation spectrum of the full chain. 
Since we know the eigenenergies and eigenvectors of 
the Hamiltonian, we can calculate dynamical quantities
such as the density of states using the Lehman representation
of the spectral function\cite{NRG_ref}.

\begin{figure}
\includegraphics[width=0.9\columnwidth]{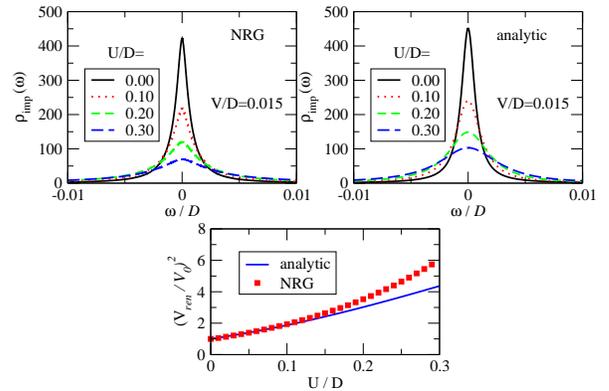}
\caption{(Color online)
Upper panels: impurity density of states for
$V/D=0.015$ and $U/D=0,\dots,0.3$ as obtained by perturbative RG and
Wilson's NRG. The lower panel shows the 
renormalized value of the 
hybridization, $V_{\rm ren}$ as a function
of the interaction strength $U$. The numerical data 
supports
the
weak coupling results for $U/D\leq0.2$.
}
\label{fig:sp_func}
\end{figure}

First let us focus on the physically relevant case of
$U>0$ and $N=2$.
To compare the numerical data with the weak coupling results, we have
calculated the impurity density of states for different values of the
interaction strength $U$. 
The results are shown in 
Fig.~\ref{fig:sp_func}. The numerical data validates the
weak coupling results for $U/D\leq0.3$. 
In our NRG calculation we considered a flat band with constant density of
states $\varrho_0=1/2D$, where $D$ stands for the half bandwidth.
In the lower panel of 
Fig.~\ref{fig:sp_func} the renormalized value of the 
hybridization, $V_{\rm ren}$ is shown as a function
of the interaction strength $U$. In NRG calculation, we have defined
$V_{\rm ren}$ from the finite size spectrum directly.
The finite size spectrum as a function of iteration number
crosses over from the initial fixed point to the strong coupling one
characterized by single particle phase shifts $\delta=\pi/2$ at around
$M^*$. $M^*$ is determined by the renormalized hybridization,
$\Delta_{\rm ren}=\pi V_{\rm ren}^2\varrho_0\sim\Lambda^{-M^*/2}$.
We take $M=M^*$ when the energy of the first excited state exceeds
$90\%$ of its fixed point value.

\begin{figure}
\vskip0.3cm
\includegraphics[width=0.8\columnwidth]{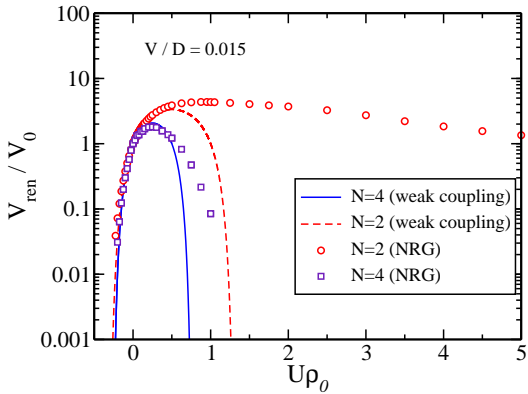}
\vskip0.3cm
\caption{(Color online)
Comparison between weak coupling RG and NRG approach:
The 
renormalized value of the 
hybridization, $V_{\rm ren}$ as a function
of the interaction strength $U$ for $N=2$ and $N=4$.
}
\label{fig:v_ren_N2N4}
\end{figure}

To answer the question whether 
an
intermediate maximum
appears outside the weak
coupling limit or not, 
we have performed calculation with
very large values of the interaction strength up to $U/D=5.0$.
The results are shown in Fig.~\ref{fig:v_ren_N2N4}.

Our conclusion is that even for $N=2$ 
such a non-monotonic behavior is found
but the position of the maximum
as well as the shape of the curve for large $U$ differs essentially
from those obtained by weak coupling calculation.
It still remains a question whether for case of many channels 
the weak coupling calculation is reliable or not.
To treat many channels with NRG is very challenging but to see the
tendency with increasing channel number, 
we performed the numerical analysis of the case $N=4$. The results
are plotted in  Fig.~\ref{fig:v_ren_N2N4} as well. Our data suggests
that already for $N=4$ the position of the turning point as well as
the decay of the curve at large $U$ is reproduced by the
weak coupling calculation with a much better accuracy than in case of
$N=2$.

\section{Anderson-Yuval approach}
\label{app:karcsi} 
\begin{figure}
\vskip0.5cm
\includegraphics[width=0.9\columnwidth]{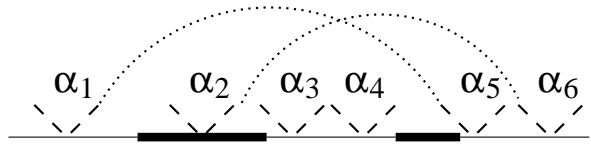}
\vskip0.5cm
\caption{The solid lines represents the time 
interval when the impurity level is occupied and the light ones stand for
unoccupied levels. The conduction electrons can join the time line
at the end points of the intervals where hopping $V_\alpha$ takes place
while they can touch the time line at any points due to the Coulomb
interaction. Those are indicated by dashed lines which are 
labeled
according to the channel indices $\alpha_i$. 
}
\label{fig:time_intvals}
\end{figure}

In most of the physical cases the Coulomb interaction $U$ dominates over the 
hopping term $V_0$. To overcome that difficulty, 
Anderson and Yuval\cite{yuval} introduced a path integral method for the
Kondo problem where the interaction $U$ is described in terms of phase
shifts while the hopping is treated as perturbation. The similarity
between the Kondo and the present problem can be exploited in the 
following way: The complex time axis is 
divided into intervals and as it is shown
in Fig.~\ref{fig:time_intvals} where the solid lines represents the time 
interval when the impurity level is occupied and the light ones stand for
unoccupied level. The conduction electrons can join the time line
at the end points of the intervals where hopping $V_0$ takes place
while they can touch the time line at any other points due to the Coulomb
interaction. Those are indicated by dashed lines which are 
labeled
according to the channel indices $\alpha_i$. Thus the incoming and 
outgoing conduction electron lines should be connected all possible ways
and finally a summation over all possible configuration of channel
labellings $\alpha_i$ must be carried out. 
Some of the connections are indicated in Fig.~\ref{fig:time_intvals}
by dotted lines.
The final result
for the partition function can be given in analytical form as in
Refs.\cite{yuval,vzz} 
where the Kondo or the two level system problems were treated.
The partition function has the form of a one dimensional Coulomb gas
with appropriately defined vector charges $C_\alpha$ ($\alpha=1\dots N-1$).
The scaling equations are derived by eliminating short time intervals
at the short time cutoff $\tau$
with its initial value taken as the inverse bandwidth
$\tau_0\sim D^{-1}$. The interaction
$V_0$ must be also made dimensionless by a factor $\sim{\tau}^{1/2}_0$.

The phase shifts for electrons in case of 
filled and empty impurity levels are:
$\delta_\alpha=-\arctan(\varrho_0\pi U)$ 
and $\delta'_\alpha=-\arctan(\varrho_0\pi U')$,
respectively.
Only their difference will appear in the scaling:
$z_\alpha=(\delta_\alpha-\delta'_\alpha)/\pi$.
The phase shifts are limited:
$|\delta_\alpha|$, $|\delta'_\alpha|<\pi/2$.
The Friedel sum rule requires $\sum_\alpha z_\alpha=-1$
which expresses that the one electron difference between
the filled and empty sites must be screened by charge oscillations
formed by the conduction electrons. The hybridization can be 
associated with the fugacity
\begin{equation}
y=V(\varrho_0\tau_0)^{1/2}\cos\delta_0\;.
\label{eq:fugacity}
\end{equation}
The interaction can be represented by charges at the interaction
points as
\begin{equation}
C_\alpha^{\pm}=\pm(z_\alpha+\delta_{\alpha0})\;
\label{eq:charge}
\end{equation}
where the index $\pm$ labels hopping in and out of the impurity level
assisted by an electron annihilation or creation in channel~0.

The lengthy derivation of the scaling equations closely follows
Refs.~\onlinecite{vzz,karcsi}) and the final result is
\begin{eqnarray}
\frac{dy}{d\ln\tau}&=&y\left(1-\frac{1}{2}\sum_\alpha C_\alpha^2\right)
\nonumber\\
\frac{d C_\alpha}{d\ln\tau}&=&-2y^2 C_\alpha\;,
\label{eq:sca_eqs_yC}
\end{eqnarray}
or expressed in terms of phase $z$ ( phase shifts) they are
\begin{eqnarray}
\frac{dy}{d\ln\tau}&=&y\left(\frac{1}{2}-z_0-\frac{1}{2}\sum_\alpha z_\alpha^2\right)
\nonumber\\
\frac{d z_\alpha}{d\ln\tau}&=&2(\delta_\alpha0+z_\alpha)y^2\;.
\label{eq:sca_eqs_yz}
\end{eqnarray}
Here the scaling is carried out by increasing $\tau$ to reduce the
electronic bandwidth. 
The term $\frac{1}{2}y$ in the first line of Eq.(\ref{eq:sca_eqs_yz})
originates in the explicite factor $\tau_0^{1/2}$ in the definition of
$y$ (\ref{eq:fugacity}) and disappears from the corresponding scaling 
equation of $V$.

The system of equations Eqs.~(\ref{eq:sca_eqs_yz}) must be solved
for initial values $y(\tau_0)\ll1$ but $z_\alpha$ can be arbitrary.
The fugacity can either increase or decrease exponentially depending
on the quantity $(z_0+\frac{1}{2}\sum_\alpha z_\alpha^2)$. Similar expressions
were obtained in Ref.\onlinecite{3dmodel} by matching the perturbative
results with expression in terms of phase shifts.

The regions for attractive interaction and large enough repulsive one
will be treated separately. In the first case 
$(-z_0-\frac{1}{2}\sum_\alpha z_\alpha^2)<0$ ($U<0$,$z_0>0$). The solution is
$y/y_0=(\tau/\tau_0)^{-z_0-\frac{1}{2}\sum_\alpha z_\alpha^2}$
thus $y$ is decreasing, therefore $z_\alpha$ ($\alpha=0,\dots$) are slowly
varying 
thus the
$\tau$ dependence 
can be
ignored. Thus
\begin{equation}
V_{\rm inv}=V_0\left(\frac{\tau}{\tau_0}\right)^{-z_0-\frac{1}{2}\sum_\alpha
    z_\alpha^2}\;.
\label{eq:vscal_yuval}
\end{equation}

The situation is different for repulsive interaction ($z_0<0$).
There are two regions in that case. In the first one 
$(-z_0-\frac{1}{2}\sum_\alpha z_\alpha^2)>0$ and $y$ is increasing.
Then Eq.~(\ref{eq:vscal_yuval}) is valid as far as $y\ll1$ is
satisfied. There is, however, a crossover where 
$(-z_0-\frac{1}{2}\sum_\alpha z_\alpha^2)=0$ to the second region
where $y$ decreases again and the screening dominates.
The larger $N$ the stronger the decrease is.

The crossover between the increasing and decreasing regions is at
$$
z_0=-\frac{1}{2}\sum z_\alpha^2\;.
$$
As $z_\alpha$-s are very slightly renormalized thus the
unrenormalized values can be used and $\delta_\alpha$ is
independent of $\alpha$ ($\delta<0$). Thus the crossover is at
$z_\alpha=2/N$ and then for 
$$
\begin{array}{ccc}
N=2: & \delta=\frac{\pi}{2}: & \varrho_0 U\to\infty\\
N=4: & \delta=\frac{\pi}{4}: & \varrho_0 U=\frac{1}{\pi}\;.
\end{array}
$$
Comparing with the results of NRG in case $N=2$ the turning point is at
$\varrho_0 U\to\infty$ thus the agreement is not complete but at least it could
be argued that it is inside the accuracy of the scaling equation. The weak 
coupling scaling result is very poor as it was expected 
(see Fig.\ref{fig:sp_func}). 
For $N=4$ all the methods give very similar results.

The general solution of the scaling equations
can be searched in the form 
$C_\alpha(\tau)=(C_\alpha)_{\rm initial}\zeta(\tau)$. Then
\begin{eqnarray}
\frac{d\zeta}{d\ln\tau}&=&-2y^2\zeta\nonumber\\
\frac{dy}{d\ln\tau}&=&
\left(1-\frac{1}{2}\zeta^2\sum_\alpha C_\alpha^2\right)y\;.
\end{eqnarray}
The scaling trajectories are 
\begin{equation}
4y^2(\tau)-\zeta^2(\tau)\sum\limits_\alpha C_\alpha^2+4\ln\zeta(\tau)=
4y^2(0)-\sum\limits_\alpha C_\alpha^2\;.
\label{eq:trajectory}
\end{equation}
During the renormalization
$y$ is fast increasing 
and the scaling is stopped where it reaches
unity. Meanwhile $\zeta$ decreases 
slowly 
and its
renormalized value 
can be extracted from 
Eq.(\ref{eq:trajectory}). The result in leading order in $y$ reads as
\begin{equation}
\ln\zeta(\tau)=-\frac{y^2(\tau)-y_0^2}{\frac{1}{2}-z_0-\frac{1}{2}\sum_\alpha z_\alpha^2}\;.
\label{eq:zetaren}
\end{equation}
Outside that region the long time approximation for the
conduction electron Green's function cannot be applied.
\section{Weak coupling approach for out of equilibrium}
\label{sec:out_of_equ}
Considering theoretical methods two ways can be followed: the Keldysh
Green's function method or the calculation of the scattering amplitude
by time-ordered perturbation theory. Here the second method will be followed,
where the initial conduction electron states can be arbitrary non-equilibrium
states and for the intermediate and final states the actual non-equilibrium
distributions are taken into account. That method has been earlier applied
in the leading logarithmic approximation\cite{fredi_lesarc,jens,ujz}, which 
is a generalization of
Anderson's poor man scaling\cite{anderson}.

Here the extension of that method is presented to next leading order. For
equilibrium first the Kondo model was treated by that way
\cite{solyom-zawadowski}. The basic idea is to calculate the 
development of the initial $|i\rangle$ state to the final
$|f\rangle$ one, but in second order the renormalization of the norms of
those states must be corrected also. Thus
the scattering matrix element to be considered is
\begin{widetext}
\begin{equation}
T_{fi}'(\omega)=\frac{\langle f|H_{\rm int}
\sum\limits_{n=0}^{\infty}\left(\frac{1}{\omega-H_0}H_{\rm int}\right)^n
|i\rangle}
{[\langle f|
\sum\limits_{n=0}^{\infty}\left(\frac{1}{\omega-H_0}H_{\rm int}\right)^n
|f\rangle
\langle i|
\sum\limits_{n=0}^{\infty}\left(\frac{1}{\omega-H_0}H_{\rm int}\right)^n
|i\rangle]^{1/2}}\;,
\label{eq:state_norm}
\end{equation}
\end{widetext}
where $\omega$ is the initial energy of state $|i\rangle$ and the Hamiltonian
is split as $H=H_0+H_{\rm int}$. 
In the present problem the correction of the
normalization occurs for the impurity electron states while in the Kondo model
for the spin states. Those normalizations appeared in the previous treatment
as $d^{1/2}(\omega)$ in Eq.(\ref{eq:v_inv}).

\begin{figure}
\includegraphics[width=0.9\columnwidth]{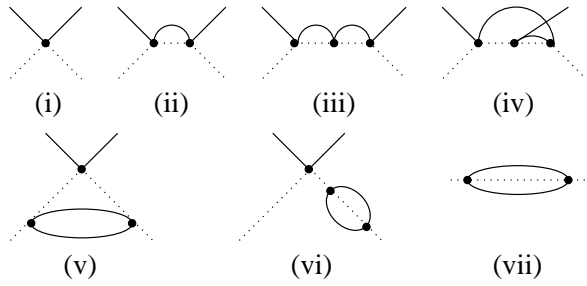}
\caption{Panels (i)-(vi): The 
diagrams
up to $\sim U^3$ order
contributing to the numerator of Eq.(\ref{eq:state_norm}). 
The diagrams should be decorated by
the direction of the lines in all possible ways. The diagram of the 
self-energy is shown in panel (vii).
}
\label{fig:Ucorr}
\end{figure}

The scaling equation can be obtained after changing the cutoff
$D\to D-\delta D$ by adjusting the coupling constant to keep
$T_{fi}$ invariant for appropriate $|i\rangle$ and $|f\rangle$.
In the following we use the original left and right states
($\alpha=L,R$) with $V_L=V_R$ as a start.
In order to derive the scaling for coupling $U$ the initial and final states
should be $a_k^\dagger d^\dagger|0\rangle$ where $|0\rangle$ is the 
non-equilibrium state at applied voltage $eV$, for which disregarding $V$ the
state is the noninteracting ground state. Considering the occupation 
$n_d$
of the
$d$-level the occupation probability in the steady state must be determined in
presence of $eV$. 
That value will be 
$\langle n_d\rangle=1/2$
for $\varepsilon_d=0$
but in the general case $\varepsilon_d\neq0$
it can depend on $eV$.
The 
diagrams of the numerator 
up to $\sim U^3$ order are shown in 
Fig.\ref{fig:Ucorr} (i)-(vi) where the diagrams should be decorated by
the direction of the lines in all possible ways. The diagram of the 
self-energy is shown in Fig.\ref{fig:Ucorr}(vii). 
In logarithmic approximation only the diagrams linear in
$\log\frac{D}{\omega}$ are contributing to the scaling equations
thus the relevant vertex corrections are (ii),(v) and (vi) while
(iii) and (iv) are not as these provide $\log^2\frac{D}{\omega}$.
The type (ii) diagrams with the parallel and antiparallel lines
cancel each other.
As the logarithmic terms in diagrams (v),(vi) and (vii) come from
closed electron loops which are independent of the actual value of
$\mu_L$ and $\mu_R$ thus to the logarithmic term the left and right
contacts contribute separately which should be independent of the
applied voltage. That simplification is not sustained in higher order 
contributions where the left and right lines 
simultaneously occur.
The self-energy
correction in (vi) contributes by adding that to either of the 
incoming and outgoing $d$-lines. One of those corrections is cancelled by the
denominator in Eq.(\ref{eq:state_norm}). As it is well known from the 
spinless fermionic case e.g. the X-ray absorption problem 
\cite{nozieres,solyom} the remaining diagram is cancelled by (v).
Thus single logarithmic term does not remain. That is similar to the
equilibrium case, see Eq.(\ref{eq:uinv_noscaling}) thus the invariant
coupling $U_{\rm inv}={\rm const}$. (For the details see Appendix A.)

In the following the renormalization of the hybridization depicted in
Fig.\ref{fig:vertex_corr}b and \ref{fig:v_corr} is crucial where e.g.
$|i\rangle=d^\dagger |0\rangle$ and  
$|f\rangle=a_{k\alpha}^\dagger |0\rangle$. Keeping terms up to 
$\sim VU^2$. After taking the
denominator in Eq.(\ref{eq:state_norm}) into account the final form
of $T_{fi}$ is
\begin{widetext}
\begin{equation}
\langle0|a_{k\alpha}Td^\dagger|0\rangle=
V_{\alpha}\left(1+U\varrho_0\log\frac{D-\varepsilon_d}{\omega-\alpha\frac{eV}{2}-\varepsilon_d}-\frac{1}{2}U^2\varrho_0^2\log\frac{D}{\omega-\varepsilon_d}\right)\;,
\label{eq:T_form}
\end{equation}
\end{widetext}
where the first correction is due to the vertex depicted in
Fig.\ref{fig:v_corr}, while the second one arises from the 
self-energy on the leg of diagram reduced by the denominator by a factor
1/2. This result agrees with Eq.(\ref{eq:v_corr2}) ($N=2$)
at $eV=0$. 
Here taking the
special case $\varepsilon_d=0$ the voltage $eV$ serves as a low energy
cutoff.

As it has been mentioned earlier considering the $d$-level there is a steady
state occupation $n_d$. That value is determined from the ballance of the 
in and outflow of the conduction electrons. To determine it 
for $\varepsilon_d\neq0$
two other
quantities must be known, namely the changes in the level position and the
spectral function of the $d$-level, $\tilde{\varepsilon}_d$ and 
$\varrho_d(\varepsilon)$ due to the applied voltage. That calculation
can be carried out numerically in a self-consistent way.

The probability of scattering of an electron coming from the left (L) or 
right (R) into the $d$-level is denoted by $W^+_{L/R}$ while the opposite
process by $W^-_{L/R}$. These quantities are
\begin{equation}
W_{L/R}^+=(1-n_d)2\pi\varrho_0\int
V_{L/R}^2(\varepsilon)\varrho_d(\varepsilon)
f_{L/R}(\varepsilon)d\varepsilon\;,
\label{eq:W+}
\end{equation}
and
\begin{equation}
W_{L/R}^-=n_d2\pi\varrho_0\int
V_{L/R}^2(\varepsilon)\varrho_d(\varepsilon)
(1-f_{L/R}(\varepsilon))d\varepsilon\;,
\label{eq:W-}
\end{equation}
where $V_{L/R}(\varepsilon)$ are determined from renormalization
group equations with the appropriate infrared cutoffs and 
$f_{L/R}(\varepsilon)=f(\varepsilon\pm eV/2)$ is the Fermi distribution
function for the leads in presence of the voltage. Those will be taken at zero
temperature $T=0$.

The steady state is determined by 
\begin{equation}
\frac{d}{dt}n_d=W_L^++W_R^+-W_L^--W_R^-=0\;.
\label{eq:steadystate}
\end{equation}
This equation combined with Eqs.(\ref{eq:W+})(\ref{eq:W-}) gives
\begin{widetext}
\begin{equation}
n_d=
\frac
{\int\varrho_d(\varepsilon)\left[V_{L}^2(\varepsilon)f(\varepsilon+eV/2)+
 V_{R}^2(\varepsilon)f(\varepsilon-eV/2)\right]d\varepsilon}
{\int\varrho_d(\varepsilon)\left[V_{L}^2(\varepsilon)+
 V_{R}^2(\varepsilon)\right]d\varepsilon}\;.
\label{eq:nd}
\end{equation}
\end{widetext}
If electron-hole symmetry holds thus $\varepsilon_d=0$, then
$n_d=1/2$.
The next step is to determine the self-energy of the $d$-electron.
The $d$-electron propagator is 
$$
\langle0|d H_1\frac{1}{\omega-H_0}H_1d^\dagger|0\rangle
$$
which can be simply developed because the occupied $d$-level determines 
the time flow. The self-energy corrections appear also in the
normalization. The effect of hybridization is just to give an extra
broadening of the $d$-level to be considered later. Without hybridization the
self-energy 
is
\begin{widetext}
\begin{equation}
\Sigma(\omega+i\delta)=
U^2\varrho_0^2\int\limits_{-D+\mu}^{D+\mu}d\xi'\int\limits_{-D+\mu}^{D+\mu}d\xi''
(1-f(\xi''))f(\xi')\frac{1}{\omega+\xi'-\xi''-\varepsilon_d+i\delta}
\label{eq:sigma}
\end{equation}
\end{widetext}
which can be evaluated as 
\begin{equation}
{\rm Re}\Sigma(\omega)=U^2\varrho_0^2\left(
|\omega-\varepsilon_d|\log\frac{|\omega-\varepsilon_d|}{D}
+2D\log2\right)\;,
\label{eq:Resigma}
\end{equation}
and
\begin{equation}
{\rm Im}\Sigma(\omega)=\pi U^2\varrho_0^2(\omega-\varepsilon_d)
\Theta(\omega-\varepsilon_d)\;.
\label{eq:Imsigma}
\end{equation}
In the equilibrium calculation the term proportional to
$\sim\omega$ is contributing to the function $d(\omega)$ in
Eqs.(\ref{eq:Gamma_d}) 
and the last constant term $\sim2D\log2$ is eliminated by
the applied counter term to keep $\varepsilon_d$
unrenormalized
while ${\rm Im}\Sigma(\omega)$ is a Korringa type of
relaxation. To be noted that the voltage does not occur as the
energy goes directly into the electron-hole creation of the same electrode. As
we mentioned at the end of Sec.\ref{sec:weak_coup}, that broadening is less
important at small energies. The hybridization of the $d$-level gives the
essential part of the broadening just like in the Anderson impurity model
\begin{equation}
\Gamma(\varepsilon)=2\pi\varrho_0(V_L^2(\varepsilon)+V_R^2(\varepsilon))\;,
\label{eq:Gamma_broad}
\end{equation}
where the voltage dependent hybridization strength must be used.

\begin{figure}
\vskip0.5cm
\includegraphics[width=0.8\columnwidth]{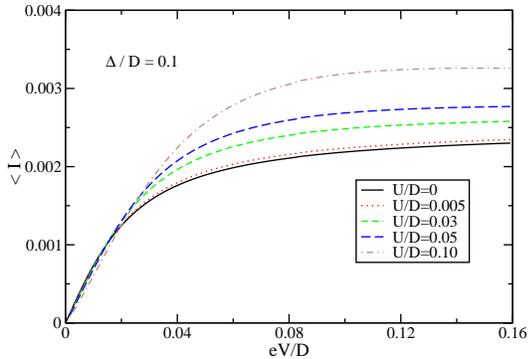}
\vskip0.5cm
\caption{(Color online)
Current obtained by weak coupling RG for
$\Delta/D=0.1$, and $U/D=0,0.005,0.03,0.05,0.10$. 
For this range of interaction strength the weak coupling
method was reliable in equilibrium.
}
\label{fig:current}
\end{figure}

The the $d$-electron spectral function is
\begin{equation}
\varrho_d(\omega)=\frac{1}{\pi}\frac{\Gamma(\omega)/2}{(\omega-\varepsilon_d)^2+(\Gamma(\omega)/2)^2}\;.
\label{eq:rho_d}
\end{equation}
With the help of these quantities we are ready to calculate the current
through th impurity:
\begin{equation}
I=\frac{1}{2}\frac{d}{dt}(N_R-N_L)=W_R^-+W_L^+-W_R^+-W_L^-.
\label{eq:curr1}
\end{equation}
Combining this equation with the expression of $W_{L/R}^{\pm}$ given in
Eqs.(\ref{eq:W+})(\ref{eq:W-}) the current takes the form
\begin{widetext}
\begin{equation}
I=n_d2\pi\varrho_0\int\left[V_R^2(\varepsilon)-V_L^2(\varepsilon)\right]
\varrho_d(\varepsilon)d\varepsilon\nonumber\\
-2\pi\varrho_0\int\left[f(\varepsilon-eV/2)V_R^2(\varepsilon)-f(\varepsilon+eV/2)
V_L^2(\varepsilon)\right]
\varrho_d(\varepsilon)d\varepsilon\;.
\label{eq:curr2}
\end{equation}
\end{widetext}
The numerical calculation goes as follows: for a fixed value of $eV$ we
discretize the energy interval $\omega_i\in[-D+\mu,D+\mu]$ and calculate
the renormalized hybridization $V_{L/R}(eV,\omega_i)$ and the 
impurity self-energy. The latter is evaluated in such a way that the
renormalized $d$-level position zero $\tilde{\varepsilon}_d=0$.
By performing a sum over $\omega_i$ we cab calculate at the
level occupation $n_d(eV)$ and the current $I(eV)$ for the given value of the
voltage. The result is shown in Fig.\ref{fig:current}
for $U/D=0,\dots,0.10$.

\begin{figure}
\vskip0.5cm
\includegraphics[width=0.8\columnwidth]{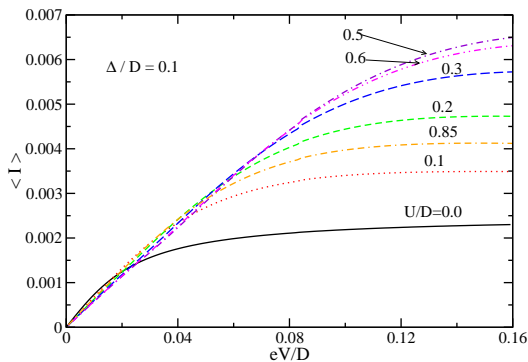}
\vskip0.5cm
\caption{(Color online)
Current obtained by the scattering formalism for
$\Delta/D=0.1$, and $U/D=0,0.10,0.20,0.30,0.50,0.60,0.85$.
Note that the weak coupling result is reliable for
$U/D\leq0.2$ only.
}
\label{fig:current_decrease}
\end{figure}

In that regime of $U$ the weak  
coupling RG gave good results in equilibrium 
(presented in Sec.~\ref{sec:nrg})
therefore one expects reliable results in out-of-equilibrium as well.
As it is shown in Fig.\ref{fig:current},
the current increases with increasing interaction
strength. The results can be interpreted as follows: 
with increasing $U$ the impurity spectral function gets broadened
since the hybridization is enhanced. 
In the linear regime ($eV<\Delta$) the $U$-dependence drops out
since the increase of coupling to the leads $\Gamma(eV,U)$ 
is cancelled by the decrease of
the height of the $d$-level spectral function which scale as
$\varrho_d(\omega=0)\sim1/\Gamma(eV,U)$. 
For larger values of the voltage $eV>\Delta$
the $d$-level is experienced not only
at the peak of the spectral function ($\varrho_0(\omega=0)$) thus
the current 
is not linear in $eV$ any more. For $eV\gg\Delta$ the complete
$d$-level contributes to the current thus it
saturates and the asymptotic value is proportional to
$\Gamma(eV,U)$.

\section{Conclusion}
\label{sec:summary}
The resonant level model studied has very different behavior for attractive
and repulsive interaction. 
This difference can be understood using the
site representation for conduction electrons in the strong $U$ limit by the
following argument:

(i) {\em in case of attractive interaction} the particle on the
$d$-level attracts electrons to pile up around 
the impurity
occupying
the next site 
thus the electron on the $d$-level is blocked
for hopping to the conduction band.

(ii) {\em in case of repulsive interaction} the site next to the
occupied $d$-level is empty thus that electron can easily jump to the
conduction band.

All the methods predict that increasing the strength of an attractive
interaction 
the hopping rate $V$ 
is reduced
and for strong enough coupling it
even goes to zero (see Fig.\ref{fig:v_ren_N2N4} for $U\varrho_0<-0.25$).
The effect of orthogonality catastrophe reducing the hopping is less
relevant
because that have been already reduced by the vertex correction.

In the repulsive case for large $U$ the orthogonality catastrophe 
(self-energy correction) is reducing essentially the hopping rate.
Thus the 
effective hopping $V_{\rm eff}$ is
first enhanced by the effect described
above and can reach a maximum which is followed by a reduction
due to the orthogonality catastrophe.
The position of the maximum can be pushed to lower energies by
increasing the number of the screening channels $N$ thus the perturbative
result becomes more and more reliable. In case of $N=2$ the latter method
leads to a pronounced maximum but the NRG indicates only slowly varying
bump. In the Anderson-Yuval approach the maximum is even pushed to
infinity. Thus the existence of the maximum is supported only by the NRG.

Considering the time ordered scattering formalism the results are in
accordance with the expectation of the weak coupling result for $N=2$.
The increasing $U$ results in increasing current as first $V$ is increased
but for larger $U$ due to the orthogonality catastrophe the current 
is essentially reduced
(see Fig.\ref{fig:current_decrease}). As the NRG does not give
a sharp crossover for the hopping rate $V_{\rm eff}$ 
thus the corresponding effect in
the current must be less pronounced in reality. 
Of course, for $N\gg2$
the crossover must clearly exist. Unfortunately, in the work of
Mehta and Andrei the crossover is in the range of reduced current, which is
just the opposite what is expected on the ground of the physical picture
established and results obtained by different methods for the hopping rate.

Very detailed further studies of the Bethe Ansatz method are needed to
understand and resolve the origin of the presented discrepancies.

\acknowledgments
The authors acknowledge the fruitful discussions with
N. Andrei, P. Mehta, 
J. von Delft, J. Kroha, P. W\"olfle, P. Schmitteckert and
G. Zar\'and. 
This work was supported by 
Projects OTKA D048665, T048782, TS049881 and
T046303. A.Z. is grateful to the Humboldt Foundation to support his stay
in Munich. L.B. acknowledges the support of the J\'anos Bolyai Scholarship.

\appendix
\section{Cancellation of the logarithmic terms in the renormalization
of the Coulomb interaction}
\begin{figure}
\includegraphics[width=0.85\columnwidth]{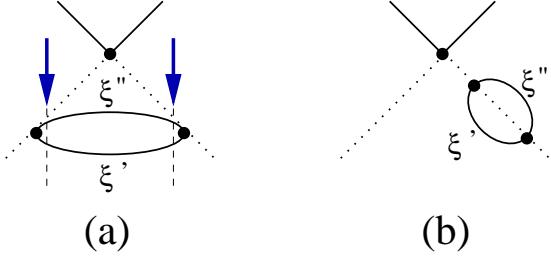}
\caption{(Color online)
The two relevant
diagrams up to $\sim U^3$ contributing to the 
numerator of Eq.(\ref{eq:state_norm}).
A Ward identity ensures the cancellation of these diagrams
for small frequencies
meaning that no logarithmic term survives in second order
thus the invariant coupling $U_{\rm inv}={\rm const}$.
}
\label{fig:app}
\end{figure}
The 
diagrams of the numerator of Eq.(\ref{eq:state_norm})
up to $\sim U^3$ order are shown in 
Fig.\ref{fig:Ucorr} (i)-(vi) and the diagram of the 
self-energy is shown in Fig.\ref{fig:Ucorr}(vii).
As noted earlier, the logarithmic terms come from 
diagrams (v),(vi) and (vii). 
Note that the self-energy
correction in (vi) 
(See also Fig.\ref{fig:app}b)
contributes by adding that to either of the 
incoming or outgoing $d$-lines. One of those corrections is cancelled by the
diagram shown in  Fig.\ref{fig:Ucorr}(vii)) in the
denominator in Eq.(\ref{eq:state_norm}). Therefore the two relevant
diagrams are those depicted in Fig.\ref{fig:app}.
The contribution of Fig.\ref{fig:app}b can be written as
\begin{widetext}
\begin{equation}
\frac{1}{\omega-\varepsilon_d}
U^2\varrho_0^2\int\limits_{-D+\mu}^{D+\mu}d\xi'\int\limits_{-D+\mu}^{D+\mu}d\xi''
(1-f(\xi''))f(\xi')\frac{1}{\omega+\xi'-\xi''-\varepsilon_d}\;.
\label{eq:appdiag1}
\end{equation}
\end{widetext}
To get the purely logarithmic term
we 
can now expand the integral in linear order in $\sim(\omega-\varepsilon_d)$
and get the form
\begin{widetext}
\begin{equation}
U^2\varrho_0^2\int\limits_{-D+\mu}^{D+\mu}d\xi'\int\limits_{-D+\mu}^{D+\mu}d\xi''
(1-f(\xi''))f(\xi')\frac{1}{(\omega+\xi'-\xi''-\varepsilon_d)^2}\;.
\label{eq:appdiag2}
\end{equation}
\end{widetext}

On the other hand, the contribution of Fig.\ref{fig:app}a
for small frequencies can be evaluated 
and the $d$-electron lines indicated by arrows occur twice
just like the denominator squared in Eq.(\ref{eq:appdiag2}).
That Ward identity ensures the cancellation of diagram
shown in Fig.\ref{fig:Ucorr}(v) by one of the leg one in
Fig.\ref{fig:Ucorr}(vi).
This means that no logarithmic term survives in second order
thus the invariant coupling $U_{\rm inv}={\rm const}$.


\begin{thebibliography}{99}
\bibitem{qdot} see e.g. L.I. Glazman, M. Pustilnik, in "Nanophysics: Coherence and Transport," eds. H. Bouchiat et al. (Elsevier, 2005), p. 427.
\bibitem{wire} see e.g. N.O. Birge and F. Pierre, in "Fundamental Problems of
  Mesoscopic Physics, Interactions and Decoherence" edited by I.V. Lerner
  B.L. Altshuler and Y. Gefen (Kluwer Academic, Dortrecht 2004) p. 3.
\bibitem{mehta}P. Mehta and N. Andrei, Phys. Rev. Lett. {\bf 96}, 216802
  (2006).
\bibitem{anderson1} P.W. Anderson, Phys. Rev. Lett. {\bf 18}, 1049 (1967).
\bibitem{resonant1} See e.g. P.B. Vigman and A.M. Finkel'stein,
Zh. Eksp. Teor. Fiz. {\bf 75}, 204 (1978) (Sov. Phys. JETP {\bf 48}, 102
(1978)). 
\bibitem{schlottmann}P. Schlottmann, Phys. Rev. B {\bf 25}, 4815 (1982).
\bibitem{3dmodel}T. Giamarchi, C.M. Varma, A.E. Ruckenstein and P. Nozi\`eres,
Phys. Rev. Lett. {\bf 70}, 3967 (1993).
\bibitem{NRG_ref}K.G. Wilson, Rev. Mod. Phys. {\bf 47}, 773 (1975);
T. Costi, in {\em Density Matrix Renormalization}, edited by
I. Peschel {\em et al.} (Springer  1999).
\bibitem{yuval}G. Yuval and P.W. Anderson, Phys. Rev. B {\bf 1}, 1522 (1970).
\bibitem{roulet} B. Roulet, F. Gavoret, P. Nozi\`eres, Phys. Rev. {\bf 178},
1072 (1969).
\bibitem{nozieres} P. Nozi\`eres and C.T. De Dominicis,
Phys. Rev. {\bf 178}, 1097 (1969). 
\bibitem{fowler}M. Fowler and A. Zawadowski, Solid State Commun. {\bf 9},
471 (1970) and A.A. Abrikosov and A.A. Migdal, J. Low Temp. Phys. {\bf 3}, 319 (1970).
\bibitem{solyom}J. S\'olyom, J. Phys F: Met. Phys. {\bf 4}, 2269 (1974).
\bibitem{vzz} K. Vlad\'ar, A. Zawadowski, and G.T. Zim\'anyi, Phys. Rev.
B {\bf 37} 2001 (1988);{\em ibid} {\bf 37} 2015 (1988).
\bibitem{karcsi} K. Vlad\'ar, Phys. Rev. B {\bf 44} 1019 (1991).
\bibitem{fredi_lesarc}A. Zawadowski in {\em Electronic Correlations:
From Meso- to Nano-Physics}, edited by T. Martin, G. Montambaux and
J. Tr\^an Thanh V\^an (EDP Sciences, Paris, 2001) p389.
\bibitem{jens}A. Rosch, J. Paaske, J. Kroha, and P. W\"olfle,
Phys. Rev. Lett. {\bf 90}, 076804 (2003); J. Paaske, A. Rosch, and
P. W\"olfle, Phys. Rev. B {\bf 69}, 155330 (2004).
\bibitem{ujz} O. \'Ujs\'aghy, A. Jakov\'ac, and A. Zawadowski,
Phys. Rev. B {\bf 72}, 205119 (2005).
\bibitem{anderson}P.W. Anderson, J. Phys. C {\bf 3}, 2436 (1970).
\bibitem{solyom-zawadowski}J. S\'olyom and A. Zawadowski, J. Phys. F:
Met. Phys. {\bf 4}, 80 (1974).
\end{thebibliography}
\end{document}